\documentstyle[12pt]{article}
\newcommand{\II}{{\cal I}}
\newcommand{\MM}{{\cal M}}
\newcommand{\NN}{{\cal N}}

\newcommand{\wb}{\bar}

\newcommand{\vo}{\vec \omega}
\newcommand{\vn}{\vec \nabla}
\newcommand{\vr}{\vec r}

\newcommand{\be}{\begin{equation}}
\newcommand{\ee}{\end{equation}}
\newcommand{\ben}{\begin{eqnarray}\displaystyle}
\newcommand{\een}{\end{eqnarray}}
\newcommand{\refb}[1]{(\ref{#1})}

\begin{document}

{}~ \hfill\vbox{\hbox{hep-th/9707123}\hbox{MRI-PHY/P970717}}\break

\vskip 3.5cm

\centerline{\large \bf A Note on Enhanced Gauge Symmetries}

\centerline{\large \bf in $M$- and String Theory}

\vspace*{6.0ex}

\centerline{\large \rm Ashoke Sen\footnote{On leave of absence from 
Tata Institute of Fundamental Research, Homi Bhabha Road, 
Bombay 400005, INDIA}
\footnote{E-mail: sen@mri.ernet.in, sen@theory.tifr.res.in}}

\vspace*{1.5ex}

\centerline{\large \it Mehta Research Institute of}
 \centerline{\large \it   Mathematics and Mathematical Physics}

\centerline{\large \it Chhatnag Road, Jhusi, Allahabad 221506, INDIA}

\vspace*{4.5ex}

\centerline {\bf Abstract}

Two different mechanisms exist in non-perturbative
String / M- theory for enhanced
$SU(N)$ ($SO(2N)$) gauge symmetries. It can appear in type IIA
string theory or $M$-theory near an $A_{N-1}$ ($D_N$) type
singularity where membrnes wrapped around two cycles become
massless, or it can appear due to coincident D-branes (and
orientifold planes) where open strings stretched between D-branes
become massless. In this paper we exhibit the relationship
between these two mechanisms by displaying a configuration in
$M$-theory, which, in one limit, can be regarded as membranes
wrapped around two cycles with $A_{N-1}$
($D_N$) type intersection matrix, and in another limit, can be
regarded as open strings stretched between $N$ Dirichlet 6-branes
(in the presence of an orientifold plane).

\vfill \eject

\baselineskip=18pt

\section{Introduction and Conclusion}

In non-perturbative string / $M$- theory, we can get enhanced gauge
symmetries at special points in the moduli space of a theory.
A particularly interesting case is that of type IIA
string / $M$- theory on a $K3$ surface, where the gauge symmetry
enhancement takes places when the $K3$ surface develops
singularities\cite{WITTD}. Typically at these 
singularities, the area of some
two cycles vanish, and hence membranes, wrapped around these two
cycles, become massless, providing the extra massless states that
are required for the gauge symmetry enhancement. The A-D-E 
classification of these
singularities is in one to one correspondence to the A-D-E
classification of enhanced gauge symmetries that arise at these
points. 
There is also an apparently unrelated mechanism for gauge symmetry
enhancement in special class of string vacua with 
Dirichlet branes and orientifold planes\cite{GIMPOL,POLC}. When $N$ of
these D-branes come on top of each other we get an $SU(N)$
enhanced gauge symmetry\cite{WITTSM}. On the other hand if 
$N$ of the D-branes
come on top of an orientifold plane, then we get an enhanced
$SO(2N)$ gauge symmetry\cite{GIMPOL}. 

The authors of ref.\cite{BERSAD} (see also \cite{VMORE})
showed that these two mechanisms
are not unrelated, but can be related to each other through a
series of $T$- and $S$-duality transformations that relates type IIA 
theory in an ALE space with $A_{N-1}$ type singularity to type IIB theory
in the background of $N$ coincident Dirichlet five-branes. In 
this paper we provide a much more direct map between these two mechanisms
by considering a configuration of Kaluza-Klein monopoles in
$M$-theory\cite{SOR,GP,RUBACK,HULLNEW,KK,BERG,IMA,KKT,PAPAD,GIBRECENT}. 
The space-time describing $N$ Kaluza-Klein monopoles in
$M$-theory has several independent two cycles whose intersection
matrix coincides with the Cartan matrix of the $A_{N-1}$ algebra. 
In the limit when all monopoles come on top of each other, each of
these two cycles acquire vanishing area, and hence membranes
wrapped around these two cycles become massless, giving rise to
enhanced gauge symmetry. On the other hand, using the
identification of $M$-theory on $S^1$ with type IIA string
theory, the configuration of $N$ Kaluza-Klein monopoles in
$M$-theory can be regarded as a configuration of $N$ parallel 
Dirichlet six
branes in type IIA string theory. We show that under this
identification, a membrane wrapped around a two cycle of the
Kaluza-Klein monopole space can be regarded as an open string
stretched between the two D6-branes in the type IIA string theory. Thus
in the language of type IIA theory, the same phenomena of
enhanced $SU(N)$ gauge symmetry can be interpreted as due to the
vanishing of the mass of the open strings stretched between the
D-branes as the D-brane positions coincide.

In type IIA string theory we can define two $Z_2$
transformations: $(-1)^{F_L}$ which changes the sign of all the
Ramond sector states on the left, and $\Omega$ which is the
world-sheet parity transformation. Whereas $(-1)^{F_L}$ generates a
symmetry of the theory by itself, $\Omega$ generates a symmetry only when
accompanied by a  parity transformation in the target space. 
If we consider type IIA string theory in ten dimensions, and take
the quotient of this theory by the $Z_2$ transformation
generated by $(-1)^{F_L}\cdot\Omega\cdot\II_3$, where $\II_3$ denotes
the reversal of sign of three of the nine space-like directions,
then the six dimensional plane, left fixed by this
transformation, is known as the orientifold plane. 
If we now consider a configuration where $N$ Dirichlet 6-branes
are on top of this orientifold plane, 
one gets an enhanced $SO(2N)$ gauge symmetry.
One can ask if this gauge symmetry enhancement can also be understood
from the $M$-theory view point. It has been shown in 
ref.\cite{SEIWITTH}
that in $M$-theory, an orientifold 6-plane of type IIA string
theory is represented by the Atiyah-Hitchin manifold\cite{AH}. 
Thus a
configuration of $N$ D6-branes and an orientifold plane will be
represented in $M$-theory by a configuration of $N$ Kaluza-Klein
monopoles and an Atiyah Hitchin space. We identify appropriate
two cycles in this space and show that their intersection matrix
coincides with the Cartan matrix of $D_N$ algebra. In the limit
when the monopoles are on top of the Atiyah-Hitchin space, the
two cycles collapse to zero size, and hence the membranes wrapped
around these two cycles become massless, giving rise to enhanced
$SO(2N)$ gauge symmetry. We show that from the point of view of
the type IIA string theory, the membranes wrapped around these
various two cycles are simply open strings stretched between
various D-branes and their images under the $Z_2$ transformation.
These become massless as the $D$-branes approach the orientifold
plane, thereby giving rise to enhanced $SO(2N)$ gauge symmetry.

In non-perturbative string theory, we also have a novel 
phenomenon involving the appearance of tensionless strings at
special points in the moduli space. This can appear in type IIB
string theory in singular background from three branes wrapped
around two cycles with vanishing area\cite{WITTTEN}, or 
in $M$-theory in the
presence of coincident five-branes, from membranes stretched
between these five-branes\cite{STROM,TOWN}. In the last
section we  study the
relationship between these two phenomena by considering type IIB
string theory in the background of Kaluza-Klein monopoles. Using
the duality between type IIB on $S^1$ and $M$-theory on $T^2$, we
can relate this to a configuration of five-branes 
in $M$-theory\cite{KKT}.  Furthermore, we show that a three
brane of type IIB string theory wrapped around a two cycle in
the Kaluza-Klein monopole space can be reinterpreted as a
membrane of $M$-theory stretched between the five-branes. Thus
when the locations of the Kaluza-Klein monopoles coincide, we get
tensionless strings in type IIB theory from three cycles wrapped
on vanishing two cycles. But the same configuration can be
reinterpreted in the $M$-theory description as membranes
stretched between coincident five-branes. This shows the
equivalence between the two mechanisms for getting tensionless
strings.

\section{Enhanced $SU(N)$ Gauge Symmetry}

The multiple Kaluza-Klein monopole solution in $M$-theory
is described by the metric:
\be \label{e1}
ds^2 = -dt^2 + \sum_{m=5}^{10} dy^m dy^m + ds_{TN}^2\, ,
\ee
where
$y^m$ denote the space-like world-volume coordinates on the
6-brane represented by this solution,  and $ds_{TN}$ is the
metric of the Euclidean multi-centered
Taub-NUT space\cite{GIBHAW}:
\be \label{e2}
ds_{TN}^2 = U^{-1} (dx^4 + \vec \omega\cdot d\vr)^2 + U d\vr^2\, .
\ee
Here $x^4$ denotes the compact direction, and
$\vr \equiv(x^1, x^2, x^3)$ 
denotes the three spatial coordinates transverse to the brane.
$U$ and $\vo$ are defined as follows:
\be \label{e3}
U = 1 + \sum_{I=1}^N {4m \over |\vr - \vr_I|}\, ,
\ee
and,
\be \label{e5}
\vn\times \vo = -\vn U\, .
\ee
$m$ and $\vec r_I$ are parameters labelling the solution.
$\vec r_I$ can be interpreted as the locations of the
Kaluza-Klein monopoles in the transverse space. 
In order that the solutions are
free from conical singularities at $\vec r=\vec r_I$, $x^4$
must have periodicity $16\pi m$.

In the multi-centered Taub-NUT space described by the
metric \refb{e2}, one can construct $N-1$ linearly independent
two cycles as follows. Consider a straight line from $\vr_i$ to $\vr_j$
in the three dimensional space labelled by $\vec r$.
{}From this we can construct a two dimensional surface in the Taub-NUT
space, by erecting at each point of this line a circle labelled by the
periodic
coordinate $x^4$. Naively, this would seem to have the geometry
of a cylinder, but if we take into account the fact that the physical
radius of the circle labelled by $x^4$ in the metric \refb{e2} is
$16\pi m U^{-1/2}$, which vanishes at $\vr=\vr_i$ and 
$\vr=\vr_j$, we see
that this surface has the topology of a sphere. Let us denote this
by $S_{ij}$.  The area of this two cycle as measured in the metric
\refb{e2}, is given by:
\be \label{e6}
\int_{S_{ij}} (U^{-1/2}(\vr)dx^4) (U^{1/2}(\vr)|d\vr|)
=16 \pi m \int_C |d\vr| = 16\pi m |\vr_i-\vr_j|\, ,
\ee
where $C$ denotes the straight line curve from $\vr_i$ to $\vr_j$
in the $\vr$ space.\footnote{Similar mass formula for three branes
wrapped on three cycles of Calabi-Yau manifolds have been derived 
previously in ref.\cite{GRV}.}
Eq.\refb{e6} also shows that if we consider a
deformation of the surface where we replace the straight line
from $\vr_i$ to $\vr_j$ by any other curve between the two
points, then the area of the surface will be proportional to the
length of this curve. Thus the surface that we have chosen,
corresponding to the straight line path, is the minimal area
surface with this topology. If $T_M$ denotes the tension of a
membrane in $M$-theory, then the mass of a membrane wrapped
around the two cycle $S_{ij}$ will be given by
\be \label{e7}
m_{ij} = 16 \pi m T_M |\vr_i-\vr_j|\, .
\ee

We can take $S_{i,i+1}$ for $1\le i\le(N-1)$ as the 
independent two cycles.  In that case the self-intersection
number for each of these cycles is 2. To see this, we
deform the surface $S_{ij}$ by replacing the straight line
path from $\vr_i$ to $\vr_j$ by any other curve between these two
points, and note that the resulting surface intersects the
original surface at two points, $\vr=\vr_i$ and $\vr=\vr_j$ (with
the same sign). $S_{i,i+1}$ intersects $S_{i-1,i}$ once (at the
point $\vr=\vr_i$) with negative sign, since for $S_{i-1,i}$ the
line is ingoing at $\vr=\vr_i$, whereas for $S_{i,i+1}$ the line
is outgoing at $\vr=\vr_i$. Finally, $S_{i,i+1}$ and $S_{j,j+1}$ 
do not intersect if $j\ne i-1, i, i+1$.
This gives the following $(N-1)\times(N-1)$
intersection matrix of the two cycles:
\be \label{e8}
I=\pmatrix{2  & -1 & 0  & 0 & \ldots & 0 & 0\cr
           -1 & 2  & -1 & 0 & \ldots & 0 & 0\cr
            0 & -1 & 2  & -1 &\ldots & 0 & 0\cr
            \cdot & \cdot & \cdot & \cdot & \ldots & \cdot &
\cdot \cr
            \cdot & \cdot & \cdot & \cdot & \ldots & \cdot &
\cdot \cr
            0 & 0  & 0  & 0  & \ldots & 2 & -1\cr
            0 & 0 &  0  & 0  & \ldots & -1 & 2\cr}\, .
\ee
This can be easily recognised as the Cartan matrix of the
$A_{N-1}$ algebra. When all the $\vr_i$'s approach each other,
the area of all the two cycles $S_{ij}$ go to zero, and we hit an
$A_{N-1}$ singularity. For $M$-theory in such a background, we
shall get extra massless states from membranes wrapped around
these two cycles, giving rise to enhanced $SU(N)$ gauge 
symmetry\cite{KKT}.

We can use the correspondence between $M$-theory on $S^1$ and
type IIA string theory to analyse how a membrane wrapped on
$S_{ij}$ is viewed in the type IIA string theory. First of all,
note that the configuration of Kaluza-Klein monopoles given in
\refb{e1} corresponds to a configuration of Dirichlet 6 branes in
type IIA theory located at $\vr_i$ ($1\le i\le N$). Since in the
identification of $M$-theory on $S^1$ with type IIA string theory
the membrane wrapped around $S^1$ (which in this case is labelled
by the coordinate $x^4$) corresponds to an elementary type IIA 
string, it is clear from the definition of $S_{ij}$
that a membrane wrapped around $S_{ij}$ will correspond to an
elementary type IIA string stretched from $\vr_i$ to $\vr_j$,
{\it i.e.} starting on the $i$th D6-brane and ending on the $j$th
D6-brane. Since the type IIA string tension $T_S$ is given by $16\pi m
T_M$ $-$ product of the membrane tension and the radius of the
compact direction (far away from the D-branes) $-$ we see that the
mass formula \refb{e7} can be rewritten as
\be \label{e9}
m_{ij} = T_S |\vr_i-\vr_j|\, 
\ee
exactly as we would expect for an open string stretched between
two D6-branes situated at $\vr_i$ and $\vr_j$. When the
$D$-branes come on top of each other, these open strings become
massless, giving rise to enhanced gauge symmetries.

This establishes the correspondence between the mechanism of
gauge symmetry enhancement in $M$-theory near an $A_{N-1}$ type
singularity, and in type IIA theory from $N$ coincident
D6-branes. In the next section we shall analyse similar
phenomenon for enhancement of gauge symmetry to $SO(2N)$.

\section{Enhanced $SO(2N)$ Gauge Symmetry}

In type IIA string theory, enhanced $SO(2N)$ gauge symmetry appears 
when $N$ D-branes coincide with an orientifold plane. 
We shall concentrate on the situation where we have
a system of $N$ D6-branes in the presence of an orientifold six
plane. As has already been mentioned in the introduction, the
orientifold six plane is obtained as the fixed point of $\II_3$
when we mod out the ten dimensional type IIA string theory by
$(-1)^{F_L}\cdot\Omega\cdot \II_3$, with $\II_3$ denoting the
transformation that reverses the sign of three spatial
coordinates (which we shall take to be $\vr = (x^1, x^2, x^3)$).

We have already seen that the D6-branes of type IIA theory are 
represented as Kaluza-Klein monopoles in $M$-theory. We shall 
now discuss how to describe an
orientifold 6-plane of type IIA theory in $M$-theory, and then
combine the two descriptions in order to describe a configuration
of orientifold plane and D-branes. This question was addressed in
ref.\cite{SEIWITTH} where it was concluded that the orientifold
six plane is
represented by the Atiyah-Hitchin space\cite{AH} 
(called $\NN$ in ref.\cite{SEIWITTH}) in $M$-theory. 
{}From far, this space looks
like $(R^3\times S^1)/\II_4$, where $\II_4$ denotes the reversal of
sign of all four coordinates labelling $R^3$ and $S^1$. Using the
correspondence between the fields in $M$-theory on $S^1$ and type
IIA string theory, it is easy to see that the transformation
$\II_4$ in $M$-theory does indeed correspond to
$(-1)^{F_L}\cdot\Omega\cdot\II_3$ in type IIA theory. Choosing a
normalization so that the asymptotic radius of $S^1$ is given by
$16\pi m$, the Atiyah-Hitchin metric can be expressed
as\cite{GIBMAN}:
\be \label{e10}
ds^2 = f(\rho)^2 dr^2 + (8m)^2 \Big( a(\rho)^2 \sigma_1^2 +
b(\rho)^2 \sigma_2^2 + c(\rho)^2 \sigma_3^2\Big)\, ,
\ee
where $f$, $a$, $b$ and $c$ are functions defined in
ref.\cite{GIBMAN}, 
\be \label{eextra}
\rho=r/8m\, ,
\ee
and,
\ben \label{e11}
\sigma_1 &=& -\sin({x^4\over 16m}) d\theta + \cos({x^4\over 16m})
\sin\theta d\phi\, , \cr
\sigma_2 &=& \cos({x^4\over 16m}) d\theta + \sin({x^4\over 16m})
\sin\theta d\phi\, , \cr
\sigma_3 &=& {1\over 16m} dx^4 + \cos\theta d\phi\, . 
\een
The coordinate ranges are given by $8m\pi \le r < \infty$,
$0\le \theta\le \pi$, $\phi$ is periodic with period $2\pi$, and
$x^4$ is periodic with period $16\pi m$. Finally, there is an
identification under the transformation $I_1$ given by:
\be \label{e12}
I_1: \quad (r, \theta, \phi, x^4) \to (r, \pi-\theta, \pi+\phi,
-x^4)\, .
\ee
It will be convenient to define a new space $\MM$, which is
described by the metric \refb{e10} before the identification by
$I_1$ given in \refb{e12}. (Note that this is different from the
double cover $\wb\NN$ discussed in \cite{SEIWITTH}, which simply
corresponds to doubling the period of $x^4$.) As shown in
\cite{GIBMAN}, the space $\MM$ has a conical singularity at
$r=8\pi m$, which gets removed when we take the quotient of this
space by $I_1$ to recover $\NN$. (Note that the complete metric
describing the orientifold 6-plane is obtained by adjoining to
\refb{e10} the (6+1) dimensional Minkowski space labelled by $t$
and $y^m$ ($5\le m\le 10$), as in eq.\refb{e1}.)

For large $r$, the metric \refb{e10} can be approximated
by (ignoring terms that are exponentially small)\cite{GIBMAN}:
\be \label{e13}
ds^2 \simeq (1 - {16m\over r})^{-1} (dx^4 + 16 m \cos\theta
d\phi)^2 + (1 - {16m\over r}) (dr^2 + r^2 d\theta^2 + r^2
\sin^2\theta d\phi^2)\, ,
\ee
which is the Euclidean Taub-NUT metric with negative mass
parameter. Comparing this with the Kaluza-Klein monopole solution
\refb{e1}, \refb{e2} we see that this solution corresponds to
$-4$ units
of magnetic charge located at $r=0$. This is consistent with the
result that the orientifold 6-plane of type IIA string theory
carries $-4$ units of 6-brane charge (in the covering space $R^3$
of $R^3/\II_3$). Upon modding out the space by $I_1$ (which
acts as $\II_3$ on $R^3$), this would
correspond to $-2$ units of six brane charge.

We shall now construct the $M$-theory background that corresponds
to $N$ D6-branes in the presence of the orientifold plane.
Unfortunately we shall not be able to write down the exact
solution, but only an approximate solution that differs from the
exact solution by terms that vanish exponentially as we go away from the
orientifold plane.  The solution is described by the metric:
\be \label{e14}
ds^2 \simeq V^{-1} (dx^4 + \vec \Omega\cdot d\vr)^2 + V d\vr^2\, ,
\ee
modded out by the transformation:
\be \label{e15}
(\vr \to -\vr, \qquad x^4\to -x^4)\, ,
\ee
where
\be \label{e16}
V = 1 - {16 m\over r} + \sum_{i=1}^N \Big( {4m\over |\vr -\vr_i|} + 
{4m \over |\vr + \vr_i|}\Big)\, ,
\ee
and, 
\be \label{e17}
\vn\times \vec\Omega = -\vn V\, .
\ee
Since $V(r)$ is invariant under $(\vr\to -\vr)$, and $\vec
\Omega\cdot d\vr$ changes sign under this transformation, the metric
\refb{e14} is invariant under \refb{e15}. Thus modding out the
space by this transformation is a meaningful concept.

Note that the metric given in \refb{e14} is singular at 
$\vr =0$ and approaches the metric given in \refb{e13}. 
This singularity is removed by replacing the metric
near $\vr=0$ 
by the Atiyah-Hitchin metric \refb{e10}, which is completely non-singular
(after we mod out by the transformation \refb{e15}, which acts as
$I_1$ given in eq.\refb{e12}). To see
that this describes an orientifold plane in the presence of $N$
D6-branes in type IIA string theory, note that near $\vr=0$ the
metric agrees with that given in \refb{e10}, and hence represents
an orientifold 6-plane of type IIA theory.
On the other hand near the point $\vr = \vr_i$ or its
image $-\vr_i$ for $1\le i\le N$, the metric agrees with the one
near a Kaluza-Klein monopole, and hence represents a D6-brane of
type IIA theory. (For $N=1$, the exact metric has been
constructed by Dancer\cite{DANCER}).

Let us now examine the two cycles in the space described by the
metric \refb{e14}. First of all there are the usual two cycles
corresponding to straight lines from $\vr_i$ to $\vr_j$ as in the
case of a configuration of Kaluza-Klein monopoles. Membranes
wrapped around these two cycles will have mass given by
eq.\refb{e7}. But now there will be additional two cycles
corresponding to straight lines joining $\vr_i$ and $-\vr_j$.
Let us denote these two cycles by $\wb S_{ij}$.
Membranes wrapped around these two cycles will have mass given
by:\footnote{Note that our derivation of the mass formula
for wrapped membranes in $M$-theory is valid only when the two
cycle does not pass close to the $\vr=0$ point, since in
computing the area of the two cycle,  we have
used the approximate metric that is valid only away from $\vr=0$.
However, we expect that the mass formula will not change even
when the correction to the metric is taken into account, since it
agrees with the exact BPS bound.} 
\be \label{e18}
\wb m_{ij} = 16 \pi m T_M |\vr_i+\vr_j|\, .
\ee
Since under the transformation \refb{e15} the two cycle $S_{ij}$
is identified with the two cycle associated with the 
line joining $-\vr_i$ and $-\vr_j$, the latter do not form an
independent set of two cycles.
We can then take the independent two
cycles to be $S_{i, i+1}$ for $(1\le i\le (N-1)$), and $\wb
S_{N-1,N}$. The intersection matrix involving $S_{ij}$'s is as
given before. $\wb S_{N-1,N}$ has self intersection number 2, as can
be seen in the same way as that for the $S_{ij}$'s. Also
$\wb S_{N-1,N}$ intersects
$S_{N-1, N}$ at two points, $\vr_{N-1}$ and $\vr_N$ (which is 
identified with its image $-\vr_N$). However, these two points
contribute with opposite sign, since the orientation of $S^1$
labelled by $x^4$ changes sign under the map \refb{e15}. Thus the
net intersection number of $S_{N-1,N}$ and $\wb S_{N-1, N}$
vanishes. Finally $S_{N-2,N-1}$ intersects $\wb S_{N-1,N}$ at
$\vr_{N-1}$ contributing $-1$ to the intersection number. Thus
the new $N\times N$ intersection matrix takes the form:
\be \label{e19}
I=\pmatrix{2  & -1 & 0  & 0 & \ldots & 0 & 0 & 0\cr
           -1 & 2  & -1 & 0 & \ldots & 0 & 0 & 0\cr
            0 & -1 & 2  & -1 &\ldots & 0 & 0 & 0\cr
            \cdot & \cdot & \cdot & \cdot & \ldots & \cdot &
\cdot & \cdot\cr
            \cdot & \cdot & \cdot & \cdot & \ldots & \cdot &
\cdot & \cdot \cr
            0 & 0  & 0  & 0  & \ldots & 2 & -1 & -1\cr
            0 & 0 &  0  & 0  & \ldots & -1 & 2 & 0\cr
            0 & 0 &  0  & 0  & \ldots & -1 & 0 & 2\cr }\, ,
\ee
where the last row and column correspond to the cycle $\wb
S_{N-1,N}$.
This can be easily recognised as the Cartan matrix of $D_N$
algebra. As all the $\vr_i$'s approach the origin, the area of
all of the cycles $S_{ij}$ and $\wb S_{ij}$ vanish, and we hit a
$D_N$ singularity.\footnote{This can also be seen explicitly from
the analysis of ref.\cite{SEIWITTH} where it was shown that the
$M$-theory background describing a configuration of $N$ D6-branes in
type IIA theory on top of an orientifold 6-plane is the space
$C^2/\Gamma_{N-2}$, where $\Gamma_{N-2}$ is the dihedral group.
This has a $D_N$ type singularity at the origin.}
For $M$-theory in such a background, the
masses of the membranes wrapped around these two cycles vanish,
giving rise to enhanced $SO(2N)$ gauge symmetry.

We shall now reinterprete this phenomenon from the viewpoint of
type IIA string theory. A repetition of the arguments for the
$A_{N-1}$ case shows that the membrane wrapped around the two
cycle $S_{ij}$ ($\wb S_{ij}$) in $M$-theory corresponds to an
open string stretched between the D6 branes situated at $\vr_i$
and $\vr_j$ (or its image at $-\vr_j$). The mass formula given in
eqs.\refb{e7}, \refb{e18} clearly reproduces the mass formula for
open strings stretched between the D6-branes (and their 
images).
Thus from the point of view of type
IIA theory, the appearance of extra massless states and hence
enhanced $SO(2N)$ gauge symmetry when the
D6-branes coincide with the orientifold 6-plane can be reinterpreted as
due to the open
strings stretched between the D6-branes and their images becoming
massless.

\section{Tensionless Strings}

Another novel phenomenon in non-perturbative string / $M$-theory
that has been discovered during the last two years is the
appearance of tensionless strings at special points in the moduli
space. This happens for example in the type IIB string theory
in a background with $A_{N-1}$ singularity. In this
case the tensionless strings come from type IIB three branes
wrapped around the two cycles of vanishing area. Another
configuration that can give rise to tensionless strings is a set
of parallel five-branes in $M$-theory. This configuration can support 
open membranes in $M$-theory, with two ends of the membrane on two
five branes. When two or more of these five-branes come close to
each other, the membranes stretched between them 
represent tensionless strings.

Following the procedure outlined in the last two sections, it is
easy to show that these two apparently different phenomena are in
fact different interpretations of the same phenomenon. 
For this we start with type IIB string
theory in the background of $N$ Kaluza-Klein monopoles. 
Using the duality between type IIB on $S^1$ and $M$-theory on
$T^2$, we can map this to a configuration of $N$ five branes in
$M$-theory on $T^2$, with the five branes being transverse to
$T^2$\cite{STROM,KKT}. In the type IIB description, we 
get string like solitonic excitations from three branes wrapped
around the two cycles $S_{ij}$. Since under the duality map
between type IIB on $S^1$ and $M$-theory on 
$T^2$\cite{SCHW,ASPIN}, a three brane
of type IIB wrapped on $S^1$ corresponds to a membrane of
$M$-theory transverse to $T^2$, we see that a three brane wrapped
on the two cycle $S_{ij}$ in type IIB theory corresponds to a membrane
stretched between the $i$th and the $j$th five-brane in
$M$-theory. When $\vr_i$ coincides with $\vr_j$, this represents
a tensionless string. This shows that the appearance of
tensionless strings in type IIB string theory
from three branes wrapped on collapsed two cycles, and
in $M$-theory from membranes stretched between coincident five
branes, are different descriptions of the same phenomenon.

\end{document}